%% file: main.tex
\documentclass[sigconf,screen]{acmart}

  \usepackage{listings}
  \usepackage{subcaption}
  \usepackage{pifont}
  \usepackage{float}
  \usepackage{cuted} 
  \usepackage{caption} 
  \lstdefinelanguage{json}{
    basicstyle=\ttfamily\small,
    showstringspaces=false,
    breaklines=true,
    literate=
     *{0}{{{\color{blue}0}}}{1}
      {1}{{{\color{blue}1}}}{1}
      {2}{{{\color{blue}2}}}{1}
      {3}{{{\color{blue}3}}}{1}
      {4}{{{\color{blue}4}}}{1}
      {5}{{{\color{blue}5}}}{1}
      {6}{{{\color{blue}6}}}{1}
      {7}{{{\color{blue}7}}}{1}
      {8}{{{\color{blue}8}}}{1}
      {9}{{{\color{blue}9}}}{1}
      {:}{{{\color{black}{:}}}}{1}
      {,}{{{\color{black}{,}}}}{1}
      {\{}{{{\color{black}{\{}}}}{1}
      {\}}{{{\color{black}{\}}}}}{1}
      {[}{{{\color{black}{[}}}}{1}
      {]}{{{\color{black}{]}}}}{1},
  }
  \lstdefinelanguage{docker}{
    keywords={FROM, RUN, COPY, CMD, WORKDIR, ENV, EXPOSE, ADD, ENTRYPOINT, ARG, LABEL, VOLUME, USER, ONBUILD, STOPSIGNAL, HEALTHCHECK, SHELL},
    keywordstyle=\color{blue}\bfseries,
    sensitive=false,
    comment=[l]{\#},
    morestring=[b]",
    morestring=[b]'
  }
  \usepackage{xcolor}
  \AtBeginDocument{%
    }

  \setcopyright{rightsretained}
  \copyrightyear{2026}
  \acmYear{2026}
  \acmDOI{10.1145/3613905.XXXXXXX}

\begin{document}
  \title{It's Not Just Timestamps: A Study on Docker Reproducibility}
  \author{Oreofe Solarin}
  \orcid{0009-0002-6083-6515}
  \affiliation{%
    \institution{Department of Computer and Data Sciences, Case Western Reserve University}
    \city{Cleveland}
    \state{Ohio}
    \country{United States}
  }
\begin{abstract}
Reproducible container builds promise a simple integrity check for software supply chains: rebuild an image from its Dockerfile and compare hashes. 
We build a Docker measurement pipeline and apply it to a stratified sample of 2{,}000 GitHub repositories that contained a Dockerfile.
We found that only 56\% produce any buildable image, and just 2.7\% of those are bitwise reproducible without any infrastructure configurations. 
After modifying infrastructure configurations, we raise bitwise reproducibility by 18.6\%, but 78.7\% of buildable Dockerfiles remain non-reproducible. 
We analyze the root causes of the remaining differences, and find that beyond timestamps and metadata, developer-controlled choices such as uncleaned caches, logs, documentation, and floating versions are dominant causes of non-reproducibility. 
We derive concrete Dockerfile guidelines from these patterns and discuss how they can inform future linters and Continuous Integration (CI) checks for reproducible containers.
\end{abstract}

  \maketitle
  \section{MOTIVATION AND PROBLEM}
  \label{sec:motivation}
  Container images have become a core shipping format for modern software, with Dockerfiles encoding how to build and package applications into OCI (Open Container Initiative) images that 
  can be pulled and deployed at scale~\cite{Bernstein2014ContainersCloud,Burns2016DesignPatternsContainer}. Large organizations increasingly deploy directly from container 
  registries rather than rebuilding from source, which means that any compromise of the build or publication pipeline can silently propagate to thousands of downstream systems, as starkly 
  illustrated by recent software supply-chain attacks such as SolarWinds~\cite{ladisa_sok_2023}. In this setting, OCI images are not just  deployment artifacts, but part of a security-critical supply chain.

  
  Reproducible builds (R-Bs) offer a way to independently rebuild an image from its
  Dockerfile and check whether the resulting image matches the published one~\cite{Lamb2022ReproducibleBuilds,benedetti_empirical_2025}.
  In practice, container reproducibility is complicated and hard to achieve because of inherited non-determinism and metadata. For example, if an independent party attempts to build any of the official Docker base images from the provided source using build instructions and metadata from the original build, it doesn't bitwise match the upstream version~\cite{osf_appendix}. 
  
  This raises a security-relevant question: \textit{how often can an independent party rebuild real-world Dockerfiles to obtain bit-identical 
  images, and what prevents it?} 
\section{RELATED WORK}
Prior work has framed build and release pipelines as a central attack surface in open-source software supply chains~\cite{ladisa_sok_2023,okafor_sok_2022}, and the R-Bs community has argued that bitwise rebuilds provide a strong integrity signal 
for consumers~\cite{fourne_its_2023,usenix_association_chainiac_2017}. Benedetti et al.\ systematically measured packaging reproducibility across package ecosystems, showing that small build variations can drastically reduce determinism and that much of the 
non-reproducibility arises from ecosystem infrastructure rather than the packages themselves~\cite{benedetti_empirical_2025}. In the Docker ecosystem, existing work primarily studies adoption, quality, and security of images and registries~\cite{LinEmpiricalStudyDocker2020,zerouali_relation_2019,shi_dr_2025}, 
or analyzes Dockerfiles and other infrastructure-as-code artifacts for smells and repair opportunities~\cite{rahman_seven_2019,henkel_shipwright_2021}. 


  \begin{figure}[!t]
  \centering
  \begin{minipage}[t]{0.48\linewidth}
  \begin{lstlisting}[language=docker,basicstyle=\ttfamily\scriptsize]
  FROM python:3.11-slim
  RUN pip install flask requests
  COPY app.py /app/
  \end{lstlisting}
  {\scriptsize Build 1: \texttt{44c6fb81...}\\Build 2: \texttt{e391a774...} \textcolor{red}{\ding{55}}}
  \par\smallskip\centering\small (a) Non-reproducible
  \end{minipage}
  \hfill
  \begin{minipage}[t]{0.48\linewidth}
  \begin{lstlisting}[language=docker,basicstyle=\ttfamily\scriptsize]
  FROM python:3.11@sha256:5be4...
  ARG SOURCE_DATE_EPOCH
  ENV SOURCE_DATE_EPOCH=$SOURCE_DATE_EPOCH
  RUN pip install flask==3.0.0
  COPY app.py /app/
  \end{lstlisting}
  {\scriptsize Build 1: \texttt{50d70b92...}\\Build 2: \texttt{50d70b92...} \textcolor{green}{\ding{51}}}
  \par\smallskip\centering\small (b) Reproducible
  \end{minipage}
  \caption{Four changes transform a non-reproducible build into a bit-identical one.}
  \label{fig:repro-fix}
  \end{figure}

  \section{STUDY DESIGN}
  \label{sec:study}
\textbf{Dataset.} 
  Using GitHub Archive, we first identify a population of popular, actively maintained repositories that modified at least one \texttt{Dockerfile} 
  in 2024. We find 19{,}262 repositories with $\geq 5$ stars that meet this criterion~\cite{osf_appendix}. From this population we draw a stratified random sample of 2{,}000 repositories using four buckets 
  based on 2024 star counts (5-7, 8-19, 20-82, $>82$ stars). Star counts serve as a rough proxy for adoption; stratification ensures that heavily starred projects are represented without 
  letting a small number of very popular repositories dominate the sample statistics.


\textbf{Rebuild Protocol.}
For each sampled repository, the pipeline attempts to: clone the repository, discover Dockerfiles, pick a single Dockerfile using a simple path-based priority (root-level Dockerfile $>$ docker/ $>$ other paths), find at least one buildable Dockerfile, and then measure reproducibility under three conditions: bitwise reproducibility, infra-reproducible, and semantic reproducibility.

\textbf{Measuring reproducibility.}
We assess the reproducibility of each buildable Dockerfile using a multi-step process.
First, we perform two clean builds of each dockerfile using the same Docker toolchain, and compute the SHA-256 
digests of the resulting images. 
If the digests are identical, we classify the Dockerfile as \emph{bitwise reproducible}. 
Else, we harden the infrastructure (set timestamps using \texttt{SOURCE\_DATE\_EPOCH}
to the Git commit timestamp and enable \texttt{rewrite-timestamp} option) to remove non-determinism from the Docker toolchain~\cite{reproducible_builds_source_date_epoch}. 
If the hardened builds now produce identical digests, we classify the Dockerfile as \emph{infra-reproducible}. 

For Dockerfiles that still remain non-reproducible, we compare per-file non-metadata contents using the \texttt{diffoci} tool~\cite{suda_diffoci}; images with no remaining semantic differences are labeled \emph{semantically reproducible}, and those with residual content 
differences are labeled \emph{non-reproducible}. 
We additionally run \texttt{diffoscope} on non-reproducible images to generate detailed reports of the differences for manual inspection~\cite{diffoscope}.
All scripts, datasets, and diff reports are available in our replication package~\cite{osf_appendix}.

\section{RESULTS}
\label{sec:results}
Our goal (RQ1) is to understand how often real-world Dockerfiles support reproducible container builds and to what extent infrastructure hardening versus developer habits explain non-reproducibility.

\textbf{Build Outcomes.}
According to Table~\ref{tab:overview}, 56.1\% of sampled projects contained at least one buildable Dockerfile under our pipeline; the rest fail due to docker build errors, cloning errors, inaccessible repositories, or timeouts. This build failure rate is in line with other large-scale studies of GitHub Docker repositories.~\cite{li2021dockermockprebuilddetectiondockerfile,EmpiricalDockerGHCito}. 


\textbf{Reproducibility outcomes.}
Among the buildable Dockerfiles, bitwise reproducibility \textit{as-is} is rare: only a small fraction produce identical image digests under repeated clean builds with the default BuildKit configuration. After enabling our hardened infrastructure configuration, the share of bitwise 
reproducible Dockerfiles increases by roughly an 18.6\%, but the majority of buildable Dockerfiles remain non-reproducible. We observed \emph{zero} semantically reproducible cases using \texttt{diffoci}~\cite{suda_diffoci}. In other words, most images do not become reproducible once infrastructure sources of non-determinism are 
removed, suggesting that developer-controlled choices in Dockerfiles themselves play a substantial role.


\begin{table}[t]
  \centering
  \caption{Overview of build and reproducibility outcomes. Percentages are relative to the 2{,}000 sampled repositories (top) and to the 1{,}123 buildable ones (bottom).}
  \label{tab:overview}
  \begin{tabular}{lrr}
    \toprule
    Outcome & Count & \% \\
    \midrule
    Total sampled repos & 2{,}000 & 100.0 \\
    Buildable (any Dockerfile) & 1{,}123 & 56.1 \\
    Not buildable & 877 & 43.9 \\
    \midrule
    As-is bitwise reproducible & 30 & 2.7 \\
    Fixed by infra changes & 209 & 18.6 \\
    Non-reproducible & 884 & 78.7 \\
    \bottomrule
  \end{tabular}
\end{table}

\textbf{Where non-reproducibility comes from.}
To understand what remains after infrastructure-level fixes, we analyze diffoscope reports for non-reproducible build pairs~\cite{diffoscope}. Table~\ref{tab:rootcauses} groups the most common residual differences. As expected, filesystem metadata 
differences (timestamps, ownership, and permissions) are essentially universal, but they don't appear alone: 43\% of inspected images also contain divergent system logs (e.g., \texttt{/var/log/}), 37\% differ in caches or on-disk databases, and 20\% 
differ in compiled artifacts (ELF binaries, bytecode). About 13\% of cases involve application-specific outputs such as generated reports or downloaded models. These patterns suggest that, once we neutralize purely infrastructural noise, a large fraction 
of remaining non-reproducibility is driven by developer-controlled behaviors in Dockerfiles.

\begin{table}[t]
  \centering
  \caption{Major causes of non-reproducibility after infrastructure configurations (based on diffoscope reports). }
  \label{tab:rootcauses}
  \begin{tabular}{lrr}
    \toprule
    Root-cause category & Count & \% \\
    \midrule
    Timestamps \& metadata & 954 & 100.0 \\
    Formatting / file ordering & 745 & 78.1 \\
    System logs & 413 & 43.3 \\
    Caches \& databases & 351 & 36.8 \\
    Compiled artifacts & 191 & 20.0 \\
    Application-specific files & 124 & 13.0 \\
    Random / non-deterministic data & 90 & 9.4 \\
    Package-manager state & 53 & 5.6 \\
    \bottomrule
  \end{tabular}
\end{table}
  


\section{DISCUSSION AND RECOMMENDATIONS}
\label{sec:implications}
\textbf{Developer-facing recommendations.} Figure~\ref{fig:repro-fix} illustrates how small, local changes can transform a non-reproducible Dockerfile into a bit-identical one; our root-cause analysis (Table \ref{tab:rootcauses}) suggests similar patterns for improving reproducibility. 
A practical takeaway for developers is to 
 (i) install system packages in a single \texttt{RUN} with \texttt{--no-install-recommends};
 (ii) Remove package index caches after every package installation;
 (iii) pin base image and package installations to specific versions; and
 (iv) avoid baking runtime state such as \texttt{/etc/machine-id}, caches, or generated reports into the final image. 
These patterns do not guarantee reproducibility, but our measurements indicate they eliminate many of the most common differences and are natural targets for future linters or CI checks.

\textbf{Implications for Supply Chain Security.} Our findings highlight a verification gap: even with hardened infrastructure, only about one in five buildable Dockerfiles support hash-based \textit{rebuild and compare} checks, limiting the practical use of R-Bs as a routine integrity mechanism in container-based supply chains.
They also suggest a responsibility shift: deterministic build infrastructure is necessary but not sufficient, because developer-controlled choices (logs, caches, floating versions) remain a dominant source of drift.

\section{STUDY LIMITATIONS AND FUTURE WORK}
\label{sec:threats}

\paragraph{Limitations.}
We discuss two limitations.
First, \emph{version drift and attribution}: registries may change between runs, and root-cause labels from \texttt{diffoci}/\texttt{diffoscope} are approximate~\cite{diffoscope,suda_diffoci}.
Then, \emph{construct validity}, we analyze one Dockerfile per repository (path-based selection) and rebuild twice on a single CI platform and architecture, so we mainly capture same-host non-determinism.

\paragraph{Future Work:} Using our rootcause analysis (Table ~\ref{tab:rootcauses}), we will build tools that identify sources of Dockerfile non-reproducibility and recommend fixes to improve reproducibility.

\section*{Acknowledgements}
I thank Dr. William Enck and Paschal Amusuo for their valuable feedback on this work.

\clearpage
\bibliographystyle{ACM-Reference-Format}
\bibliography{FSEreferences}
\clearpage

\input{appendix.tex}

\end{document}

%% file: appendix.tex
\appendix

\lstdefinelanguage{docker}{
    keywords={FROM, RUN, CMD, LABEL, MAINTAINER, EXPOSE, ENV, ADD, COPY, ENTRYPOINT, VOLUME, USER, WORKDIR, ARG, ONBUILD, STOPSIGNAL, HEALTHCHECK, SHELL, AS},
    keywordstyle=\color{blue}\bfseries,
    identifierstyle=\color{black},
    sensitive=false,
    comment=[l]{\#},
    commentstyle=\color{gray}\ttfamily,
    stringstyle=\color{red}\ttfamily,
    morestring=[b]',
    morestring=[b]",
    basicstyle=\ttfamily\small,
    breaklines=true,
    showstringspaces=false,
}

\section{Experiment Listings and Artifacts}
\label{appendix:listings}

This appendix provides concrete examples of non-reproducibility patterns observed in our study, including actual Dockerfile listings, diffoscope output, and layer-by-layer comparison data.

\subsection{Official Docker Hub Image Comparison}
\label{appendix:official-image}

We attempted to rebuild the official \texttt{golang:1.25.5-alpine3.22} image from its source Dockerfile and compared it against the image published on Docker Hub. Despite using the exact same Dockerfile and build context from the official repository, the resulting image differed.

\subsubsection{Layer-by-Layer Comparison}

\begin{table}[h]
\centering
\small
\caption{Layer Comparison: Local Build vs Docker Hub (golang:1.25.5-alpine3.22)}
\label{tab:golang-layers}
\begin{tabular}{c|c|c|c}
\hline
\textbf{Layer} & \textbf{Local SHA256} & \textbf{Docker Hub SHA256} & \textbf{Match} \\
\hline
1 (Alpine base) & \texttt{2d35ebdb...} & \texttt{2d35ebdb...} & \ding{51} \\
2 (Go env setup) & \texttt{a8f1b7b4...} & \texttt{2cca2393...} & \ding{55} \\
3 (Go SDK) & \texttt{5c445a0e...} & \texttt{5c445a0e...} & \ding{51} \\
4 (WORKDIR) & \texttt{a9038427...} & \texttt{09615bee...} & \ding{55} \\
5 (Empty) & \texttt{4f4fb700...} & \texttt{4f4fb700...} & \ding{51} \\
\hline
\end{tabular}
\end{table}

Despite having identical source code and Dockerfile, 2 out of 5 layers differ. The differences are attributed to:
\begin{itemize}
    \item \textbf{Layer 2}: Environment variable setup creates files with different timestamps (291,165 bytes vs 291,156 bytes)
    \item \textbf{Layer 4}: WORKDIR creation has metadata differences (124 bytes vs 126 bytes)
\end{itemize}

\subsubsection{OCI Manifest Comparison}

\begin{lstlisting}[language=json,caption={Local Build Manifest (abbreviated)},label=lst:local-manifest]
{
  "config": {
    "digest": "sha256:a021bbe882fc95c51accbffd4fbe4d95...",
    "size": 2165
  },
  "layers": [
    {"digest": "sha256:2d35ebdb...", "size": 3802452},
    {"digest": "sha256:a8f1b7b4...", "size": 291165},
    {"digest": "sha256:5c445a0e...", "size": 60151314},
    {"digest": "sha256:a9038427...", "size": 124},
    {"digest": "sha256:4f4fb700...", "size": 32}
  ]
}
\end{lstlisting}

\begin{lstlisting}[language=json,caption={Docker Hub Manifest (abbreviated)},label=lst:upstream-manifest]
{
  "config": {
    "digest": "sha256:60343179d354cc763b5b4ab89f5093...",
    "size": 2171
  },
  "layers": [
    {"digest": "sha256:2d35ebdb...", "size": 3802452},
    {"digest": "sha256:2cca2393...", "size": 291156},
    {"digest": "sha256:5c445a0e...", "size": 60151314},
    {"digest": "sha256:09615bee...", "size": 126},
    {"digest": "sha256:4f4fb700...", "size": 32}
  ],
  "annotations": {
    "org.opencontainers.image.created": "2025-12-02T17:46:38Z",
    "org.opencontainers.image.source": "https://github.com/..."
  }
}
\end{lstlisting}

\textbf{Key Observation}: Even the config digest differs (2165 bytes vs 2171 bytes), indicating the image configuration contains build-time metadata that differs between builds.

\subsection{Non-Reproducibility Root Cause Taxonomy}
\label{appendix:taxonomy}

From analysis of 954 diffoscope reports, we identified the following root causes:

\begin{table}[h]
\centering
\small
\caption{Root Cause Taxonomy from Diffoscope Analysis}
\label{tab:taxonomy}
\begin{tabular}{l|r|p{3.5cm}}
\hline
\textbf{Category} & \textbf{Prevalence} & \textbf{Example Files} \\
\hline
Filesystem metadata & 100\% & timestamps, file ordering \\
File ordering & 78\% & directory listings \\
System logs & 43\% & \texttt{/var/log/dpkg.log}, \texttt{apt/term.log} \\
Cache databases & 37\% & \texttt{ldconfig/aux-cache}, \texttt{fontconfig/} \\
pip/Python & 7.7\% & \texttt{.pyc} files, \texttt{site-packages/} \\
npm/Node.js & 1.7\% & \texttt{node\_modules/}, \texttt{.npm/\_cacache/} \\
Maven/Java & 0.8\% & \texttt{.m2/}, \texttt{.class} files \\
\hline
\end{tabular}
\end{table}

\subsection{Example diffoci Output}
\label{appendix:diffoci}

\subsubsection{Python Project with Bytecode Differences}

\begin{lstlisting}[basicstyle=\tiny\ttfamily,caption={diffoci output for Bagautdino/ai-sast-tool},label=lst:diffoci-python]
TYPE  NAME                                               INPUT-0     INPUT-1
File  usr/local/lib/python3.12/site-packages/
      requests/__pycache__/_internal_utils.cpython-312.pyc
                                                         5e95576c... e6fbe4a7...
File  usr/local/lib/python3.12/site-packages/
      httpcore/_sync/__pycache__/connection_pool.cpython-312.pyc
                                                         cb00a00e... 5d3d62e9...
File  usr/local/lib/python3.12/site-packages/
      pydantic/_internal/__pycache__/_known_annotated_metadata.cpython-312.pyc
                                                         ef03b3fe... 9b375b90...
\end{lstlisting}

\textbf{Root Cause}: Python bytecode (\texttt{.pyc}) files contain timestamps by default. Solution: Use \texttt{PYTHONDONTWRITEBYTECODE=1} or compile with \texttt{SOURCE\_DATE\_EPOCH}.

\subsubsection{Node.js Project with npm Cache Differences}

\begin{lstlisting}[basicstyle=\tiny\ttfamily,caption={diffoci output for yu-yaba/gatareview-front},label=lst:diffoci-npm]
TYPE  NAME                                               INPUT-0     INPUT-1
File  root/.npm/_cacache/index-v5/cd/fd/
      75d428c6fc62506e4e534279fa1f961a698ba60154f4c154b3659f5c5182
                                                         fc0b25ac... 628d0ca8...
File  root/.npm/_cacache/index-v5/3d/dd/
      3e8c6b6c875addf2057fb5fa152fff0fab217f78a8e51c0aa457a925bf12
                                                         31e9d88c... 3cfd8c5f...
File  root/.npm/_cacache/index-v5/4e/be/
      da18b216664b1e58df4ffd6782d77f3c28a3f61355544535ff7704cc1b2d
                                                         be2e9034... 8dd0b68e...
\end{lstlisting}

\textbf{Root Cause}: npm caches package metadata with timestamps. Solution: \texttt{npm ci --cache /tmp/npm \&\& rm -rf /tmp/npm}.

\subsubsection{Debian/Ubuntu with System Logs}

\begin{lstlisting}[basicstyle=\tiny\ttfamily,caption={diffoci output for sklyar/docker-log-driver-telegram},label=lst:diffoci-apt]
TYPE  NAME                        INPUT-0         INPUT-1
File  var/log/apt/history.log     edc2b16792...   da9d2416b2...
File  var/log/dpkg.log            266bb777be...   db5ed80c6e...
File  var/log/apt/term.log        9e297133d7...   b2cc0f96ab...
\end{lstlisting}

\textbf{Root Cause}: APT logs contain installation timestamps. Solution: \texttt{rm -rf /var/log/apt/* /var/log/dpkg.log} in Dockerfile.

\subsection{Example Dockerfiles with Issues}
\label{appendix:dockerfiles}

\subsubsection{Non-Reproducible Dockerfile: Missing Cleanup}

\begin{lstlisting}[language=docker,caption={arc53/DocsGPT - npm cache not cleaned},label=lst:npm-issue]
FROM python:3.12-bookworm
RUN curl -fsSL https://deb.nodesource.com/setup_20.x | bash - \
    && apt-get install -y nodejs \
    && rm -rf /var/lib/apt/lists/*
RUN npm install -g husky vite    # PROBLEM: cache persists
\end{lstlisting}

\textbf{Issue}: \texttt{npm install -g} creates cache at \texttt{\textasciitilde/.npm/\_cacache} with timestamps.

\textbf{Fix}:
\begin{lstlisting}[language=docker]
RUN npm install -g husky vite && npm cache clean --force
\end{lstlisting}

\subsubsection{Non-Reproducible Dockerfile: pip without --no-cache-dir}

\begin{lstlisting}[language=docker,caption={abetlen/llama-cpp-python - Missing pip flags},label=lst:pip-issue]
FROM nvidia/cuda:12.5.1-devel-ubuntu22.04
RUN apt-get update && apt-get upgrade -y \
    && apt-get install -y git build-essential \
    python3 python3-pip gcc wget \
    ocl-icd-opencl-dev opencl-headers clinfo \
    libclblast-dev libopenblas-dev
    # PROBLEM: No apt cleanup, no man page removal

RUN python3 -m pip install --upgrade pip pytest cmake...
    # PROBLEM: Missing --no-cache-dir
\end{lstlisting}

\textbf{Issues}:
\begin{enumerate}
    \item apt lists not cleaned (\texttt{/var/lib/apt/lists/*})
    \item Man pages installed (\texttt{/usr/share/man/*})
    \item pip cache persists (\texttt{\textasciitilde/.cache/pip})
\end{enumerate}

\textbf{Fix}:
\begin{lstlisting}[language=docker]
RUN apt-get update && apt-get install -y --no-install-recommends \
    git build-essential python3 python3-pip gcc wget ... \
    && rm -rf /var/lib/apt/lists/* /usr/share/man /usr/share/doc
    
RUN pip install --no-cache-dir --upgrade pip pytest cmake...
\end{lstlisting}

\subsection{Reproducible Dockerfile Template}
\label{appendix:template}

Based on our findings, we recommend the following template for reproducible Dockerfiles:

\begin{lstlisting}[language=docker,caption={Template for Reproducible Dockerfile},label=lst:template]
# syntax=docker/dockerfile:1
FROM python:3.12-slim

# Reproducibility: Set epoch for timestamps
ARG SOURCE_DATE_EPOCH
ENV SOURCE_DATE_EPOCH=${SOURCE_DATE_EPOCH:-0}

# Prevent interactive prompts
ENV DEBIAN_FRONTEND=noninteractive

# Install system deps with full cleanup
RUN apt-get update && \
    apt-get install -y --no-install-recommends \
        build-essential curl && \
    # Remove caches and non-essential files
    rm -rf /var/lib/apt/lists/* \
           /var/cache/apt/archives/* \
           /usr/share/man \
           /usr/share/doc \
           /var/log/* && \
    # Reset machine-id
    truncate -s 0 /etc/machine-id

# Python deps without cache
COPY requirements.txt .
RUN pip install --no-cache-dir -r requirements.txt && \
    find /usr -type d -name __pycache__ -exec rm -rf {} + 2>/dev/null || true

# Node.js deps (if needed) with cache cleanup
COPY package*.json ./
RUN npm ci --cache /tmp/npm && \
    rm -rf /tmp/npm ~/.npm

# Application code
COPY . .

# Build command (for compiled languages)
# RUN go build -trimpath -ldflags="-buildid=" -o /app/binary .
\end{lstlisting}

\textbf{Build command for reproducibility}:
\begin{lstlisting}[language=bash]
docker build \
  --build-arg SOURCE_DATE_EPOCH=$(date +%s) \
  --provenance=false \
  --sbom=false \
  -t myimage:latest .
\end{lstlisting}